\documentclass[pre,aps,showpacs,twocolumn,floatfix]{revtex4}
\usepackage{graphicx}
\usepackage{epsfig}
\usepackage{dcolumn}
\usepackage{bm}
\usepackage{amsfonts}
\usepackage{color}
\usepackage{amssymb}

\begin{document}
\title{Reorganization of a dense granular assembly:
the `unjamming response function'}
\author{\'Evelyne Kolb, Jean Cviklinski, Jos\'{e} Lanuza, Philippe Claudin
and \'Eric Cl\'{e}ment}
\affiliation{
Laboratoire des Milieux D\'{e}sordonn\'{e}s et H\'{e}t\'{e}rog\`{e}nes
\footnote{Present address:
Laboratoire de Physique et M\'ecanique des Milieux H\'et\'erog\`enes,
ESPCI, 10 rue Vauquelin, 75231 Paris Cedex 05, France.}\\
UMR7603 -- Universit\'{e} Pierre et Marie Curie -- Bo\^{\i}te 86\\
4, Place Jussieu, 75252 Paris Cedex 05, France.}

\begin{abstract}
We investigate the mechanical properties of a static dense granular assembly
in response to a local forcing. To this end, a small cyclic displacement is
applied on a grain in the bulk of a 2D disordered packing under gravity and
the displacement fields are monitored. We evidence a dominant long range
radial response in the upper half part above the sollicitation and after a
large number of cycles the response is `quasi-reversible' with a remanent
dissipation field exhibiting long range streams and vortex-like symmetry.
\end{abstract}

\maketitle


It is always striking that apparently different cellular materials
like foams, emulsions or granular systems share in common many rheological
properties \cite{LiuNature}. All these systems can flow like fluids when a
sufficiently high external stress is applied but jam into an amorphous rigid
state below a critical yield stress. This jamming transition is associated
with a slowdown of the dynamics which led Liu et al.\cite{LiuNature} to
propose an analogy between the process of jamming and the glass transition
for glass-forming liquids. Although the nature of this jamming transition is
still unclear experimentally \cite{D'Anna,DaCruz02}, several theoretical
attempts were made to adapt the concepts of equilibrium thermodynamics to
athermal systems out of equilibrium \cite{Edwards,Ono,Coniglio}. For
packing made of grains with a size larger than few microns, thermal
fluctuations are too small to allow a free exploration of the phase space
and grains are trapped into metastable configurations. The system can not
evolve until an external mechanical perturbation like shear or vibration is
applied, which allows grains to overcome energy barriers and triggers
structural rearrangements. In this case, the free volume and configurations
accessible to each grain are capital notions that were used to define the
new concept of `effective temperature' \cite{Edwards}. Recently it was
proposed that this notion could account for the transports properties in the
vicinity of a jammed state \cite{Kurchan}. However besides this large number
of theoretical and numerical works there are only few experiments connecting
the motion at the grain level to the macroscopic mechanical behavior
\cite{Pouliquen}. It appears that there is a crucial need for understanding
the connection between the local geometrical properties and the possible
motion of a grain since structural rearrangements, and therefore displacement
fields, are the key for understanding the rheology of dense systems. 

On the other hand, there is still a debate to understand the elasto-plastic
behavior of amorphous materials \cite{Goldenberg,Gay,Wittmer,Langer}. Recent
experiments on the response of a granular pile to a small force perturbation
revealed an elastic-like behavior, which is very sensitive to the preparation,
i.e. to the microscopic texture \cite{Reydellet}. Similar conclusions were
drawn in the context of sound propagation \cite{Gilles}. However, at this
stage it is not entirely clear which features of the texture (such as
coordination number, contact distribution, various fabric tensors) are
useful to build relevant macroscopic constitutive relations \cite{Makse}.
When the external drive increases an irreversible yield occurs that is
usually described by a plasticity theory. Yet the study of the early stages
of plasticity is of crucial interest for a better understanding of yield
properties and the elucidation of strain localization (shear bands).
Recent theoretical attempts were made to explain global plastic
deformations from a modelling of local structural rearrangements named `shear
transformation zone'. This approach was firstly introduced to account
for the onset of plasticity in amorphous solids \cite{Langer} but was
extended to granular materials \cite{Lemaitre}.

In this paper, we present a conceptually simple experiment which aim is to
study the response of a dense disordered granular media to a small
perturbation induced by the displacement of a grain-scale intruder.
The forces applied to the intruder grain are large enough to unjam this
initially static packing but the driving is slow enough to stay in the
\emph{quasi-static} regime. Deformations induced by the local perturbation
are small (less than $10^{-2}$) and so we are rather far from a fully
developed plasticity regime that would be induced, for example, by a
moving rod \cite{Schiffer}. We propose a new path of study for the jammed
state by monitoring the displacement fields in response to a localized
cyclic perturbation that brings the system above the jamming transition.
We call it the `unjamming response' function which should be a
characteristic feature of the packing configurations and of its
reorganization properties. Note that, very recently, similar displacement
response experiments have been performed by Moukarzel \emph{et al.}
\cite{Moukarzel}.

The typical displacements induced by the perturbation are small at the scale
of a grain: we observe a range of displacements between $1/2000$
and $1/3$ of a grain size $d$ but these are still very large compared to the
local displacements induced by the granular contact deformations. A simple
order of magnitude calculation shows that grains can be considered as rigid,
since for the metallic grains we use and the load experienced by the packing
under gravity, elastic displacements at contacts are as small as $10^{-8}d$.
This huge separation of scales shows that we are probing the
response of the granular assembly solely due to grain reorganizations. It
corresponds either to contact opening/closing or to a change of contact
direction (rolling contact). Under gravity these processes may be partially
reversible and depend in a very sensible way on the packing geometrical
properties (texture, density) and how far we are from the jamming
transition. This sensitivity can be seen as the hallmark of the `fragile'
character of granular assemblies and the questions raised to understand this
relation are important to obtain a fully consistent picture of condensed
jammed phases as described for example by O'Hern \emph{et al.} \cite{OHern}.

\begin{figure}[t]
\begin{center}
\epsfig{file=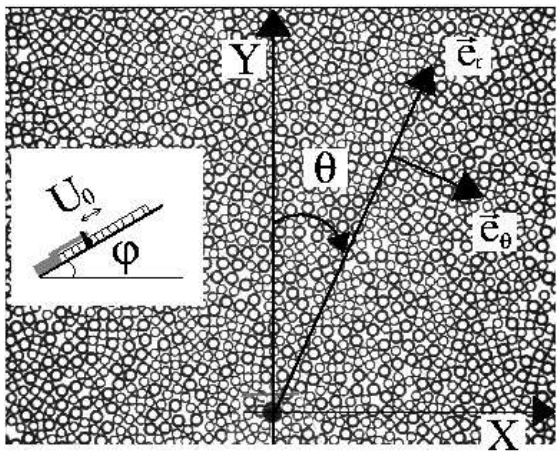,width=0.85\linewidth}
\end{center}
\caption{
Top view of the piling (inset: sketch of the experimental set up viewed
from the side).}
\label{photoecrou}
\end{figure}

A sketch of the experimental set-up is shown in figure \ref{photoecrou}, inset.
We prepare disordered $2D$ packings of brass hollow cylinders by mixing small
(diameter $d_{1}=4$ mm) and big ones ($d_{2}=5$ mm) in order to avoid
macroscopic crystallization. We chose to have the total mass of small
cylinders to be equal to that of the big ones, so that their numbers are
respectively in proportion of $7$ to $4$. This geometry allows a precise
monitoring of the displacements for each grain in the visualization field.
All cylinders have a $3$ mm height and lay on a low frictional glass plane.
The lateral and bottom walls are made of plexiglass and delimits a
rectangular frame of $L=26.8$ cm $\simeq 54 d_2$ width and an adjustable
height of typically $H=34.4$ cm $\simeq 70 d_2$. The bottom plane can be
tilted at an angle $\varphi =33^{o}$ such as to control the confinement
pressure inside the granular material by an effective gravity field
$g\sin\varphi$. The angle $\varphi$ value is larger than the static Coulomb
angle of friction between the grains and the glass plate. The intruder is
a big grain of diameter $d_2$ located in the median part of the container
at a $21.2$ cm (i.e. $\simeq 42d_2$) depth from the upper free surface.
It is attached to a rigid arm in plexiglass moved by a translation stage
and a stepping motor driven by a computer, so that the intruder motion is
characterized by cycles of displacement along the median axis $Y$ of the
container. In this report the intruder is moving up then down in a
quasi-static way at a velocity of $156$ $\mu$m/s separated by rest periods
of $9$ s. The intruder displacement value $U_{0}$ is only a fraction of
a grain diameter ($U_{0}=1.25$ mm). A high resolution CCD camera
($1280\times 1024$ pixels$^{2}$) is fixed above the experimental setup
with its plane parallel to the tilted plane which is homogeneously
illuminated from behind. The image frame is centered slightly above the
intruder and covers a zone of area $39d_{2}\times 31d_{2}$ -- see figure
\ref{photoecrou}. The camera is triggered by a signal coming from the
motor which allows the capture of an image in the rest phase one second
before each intruder displacement. In the following, we use the notation $i$
for the index corresponding to the $i^{th}$ image just before the $i^{th}$
displacement (upwards or downwards) and $n$ for the cycle number with
$n=int\left( \frac{i+1}{2}\right)$. The center of each grain is determined
with precision using the computation of the correlation between an image of
the packing and two reference images corresponding to both grain types
($d_{1}$ or $d_{2})$. We obtain, for each image, the locations of more than a
thousand grains with a resolution down to $0.05$ pixels. The displacements
of each single grain in response to the intruder motion is determined with a
precision of less than $10$ $\mu$m.

\begin{figure}[t]
\begin{center}
\epsfig{file=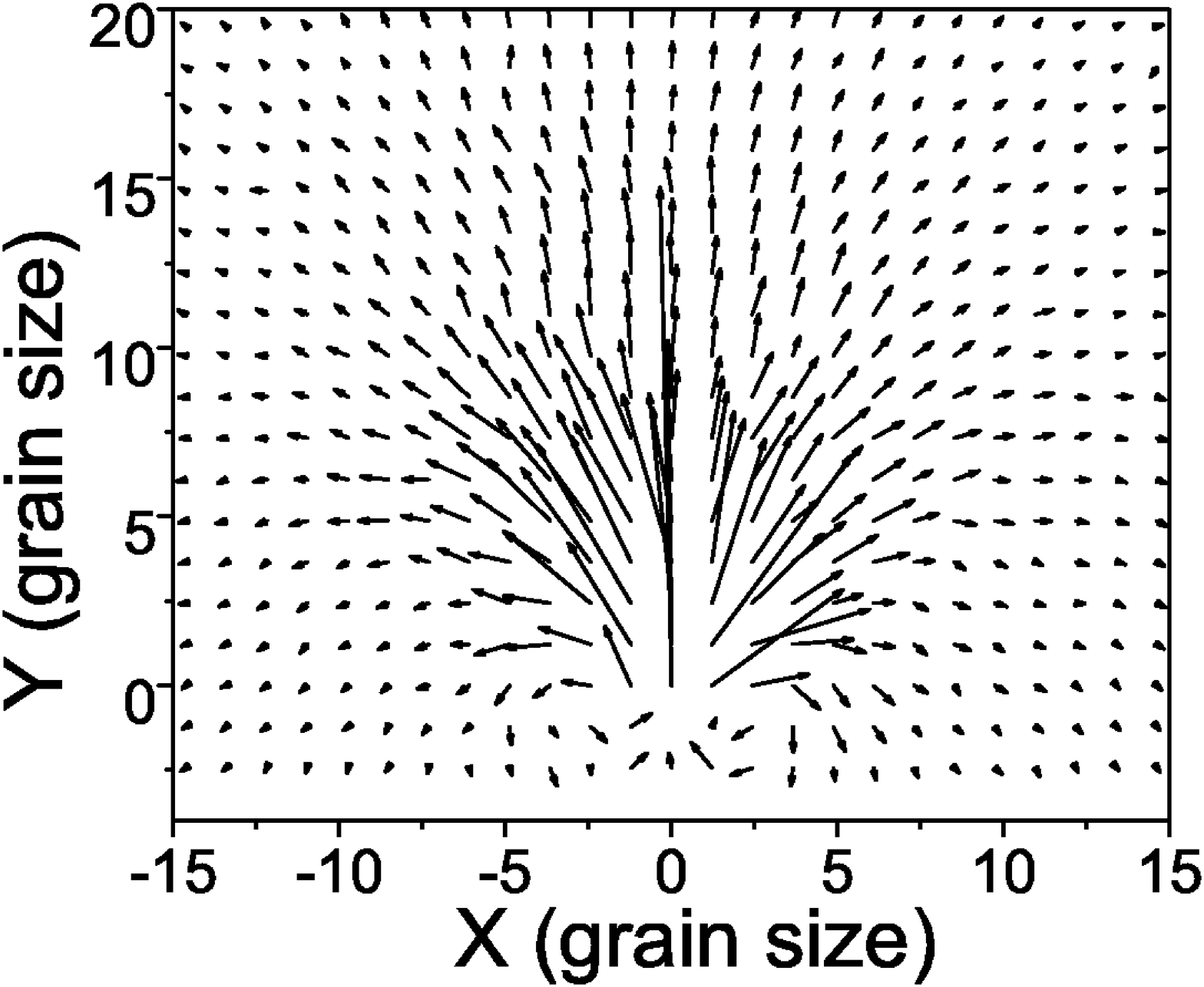,width=0.93\linewidth}
\end{center}
\caption{
Averaged displacement field for the second upward motion of the intruder
($n=2,i=3$). Note that all displacements are magnified by a factor of $70$.
The intruder is located at $X=Y=0$.}
\label{champdepl}
\end{figure}

\begin{figure*}[t]
\begin{center}
\epsfig{file=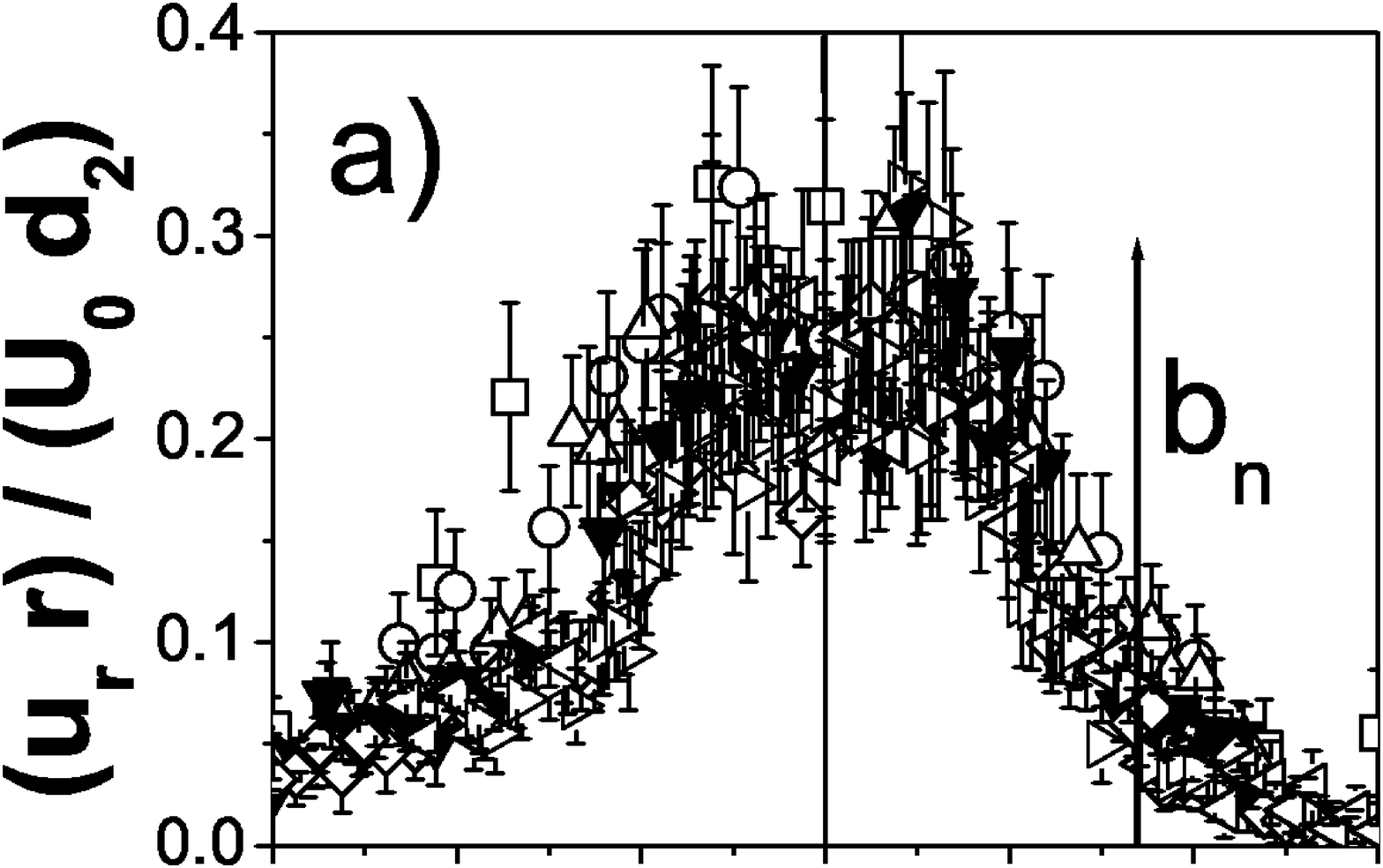,width=0.45\linewidth}
\hspace*{0.05cm}
\epsfig{file=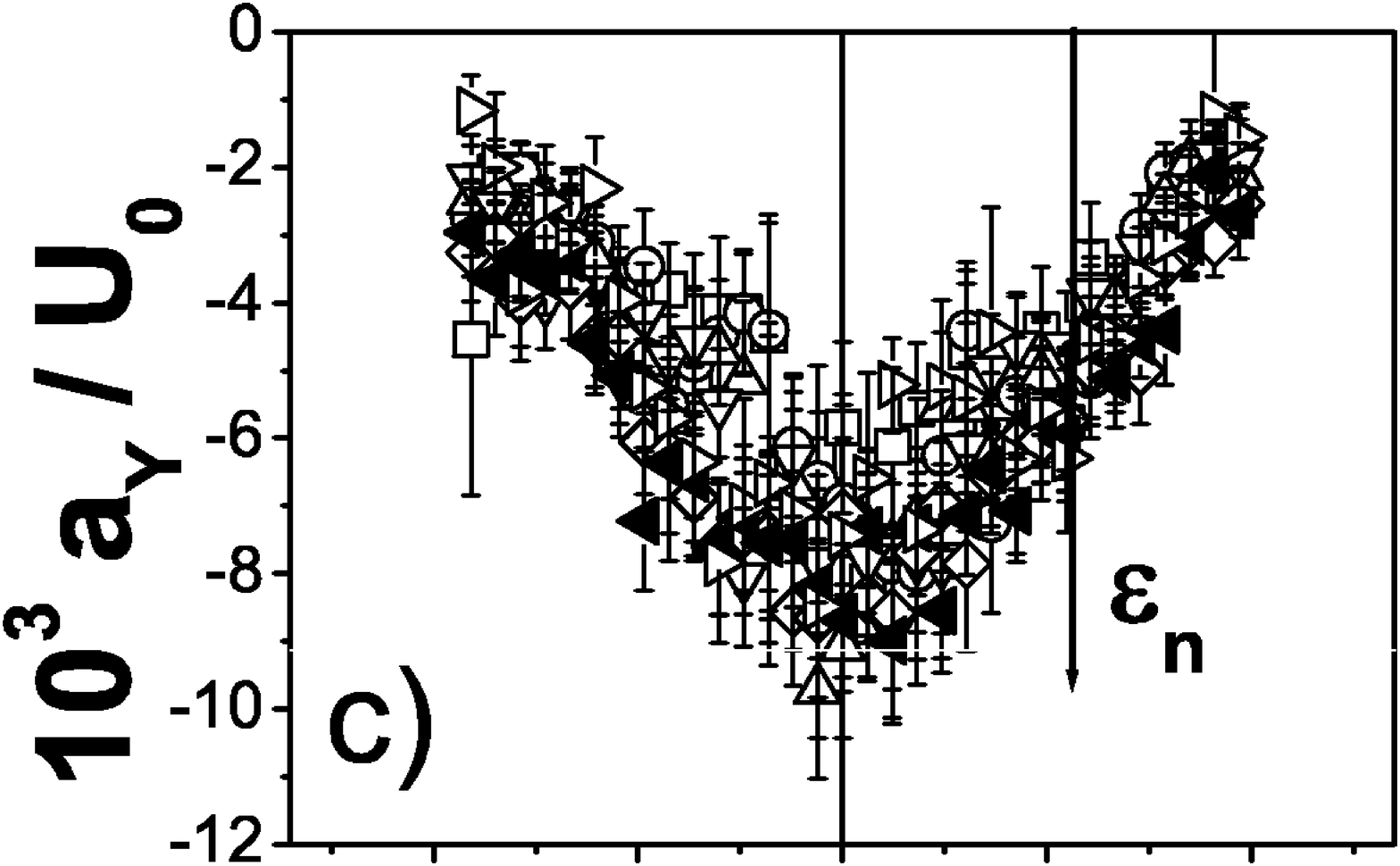,width=0.455\linewidth}
\epsfig{file=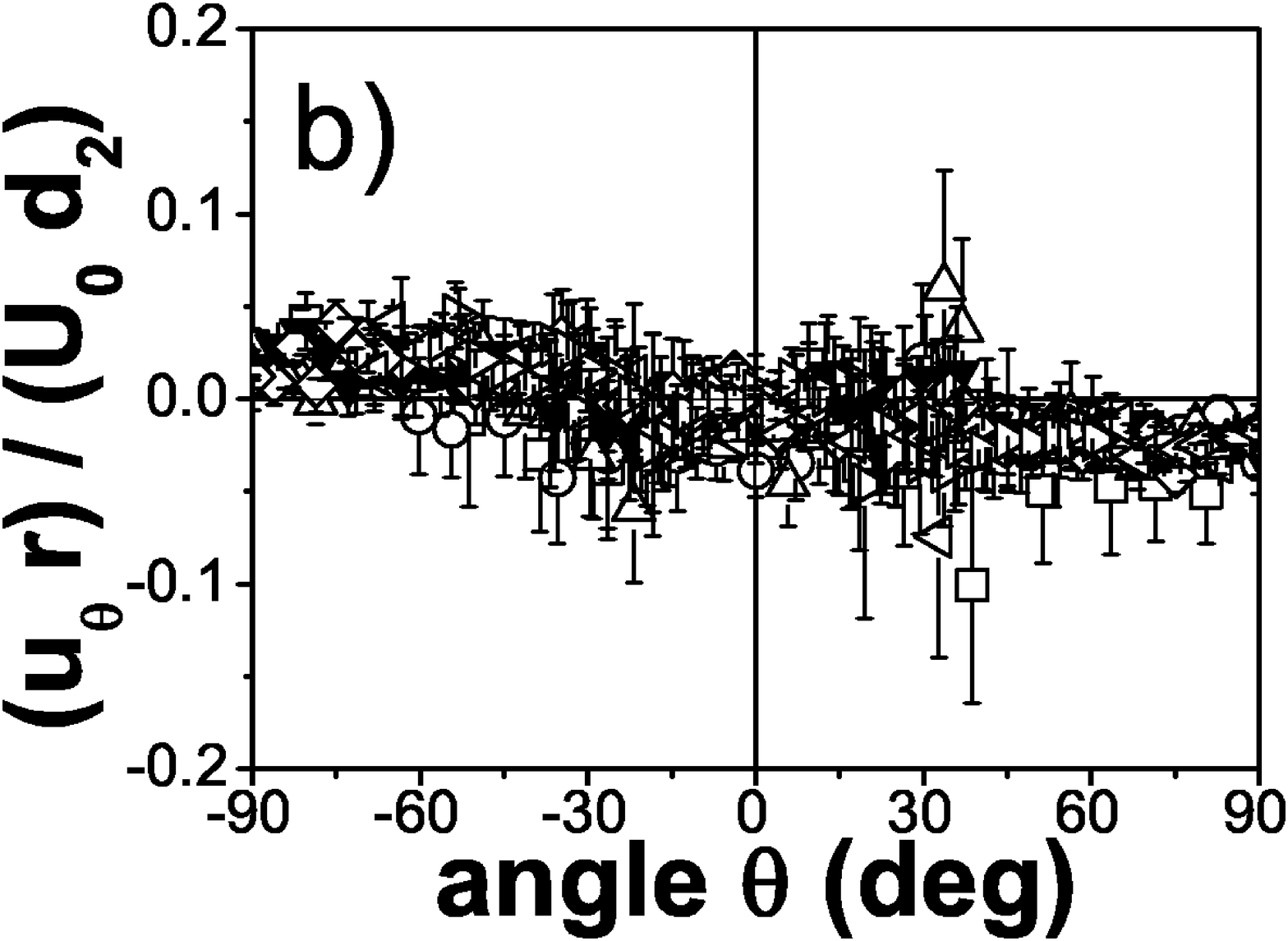,width=0.45\linewidth}
\hspace*{0.1cm}
\epsfig{file=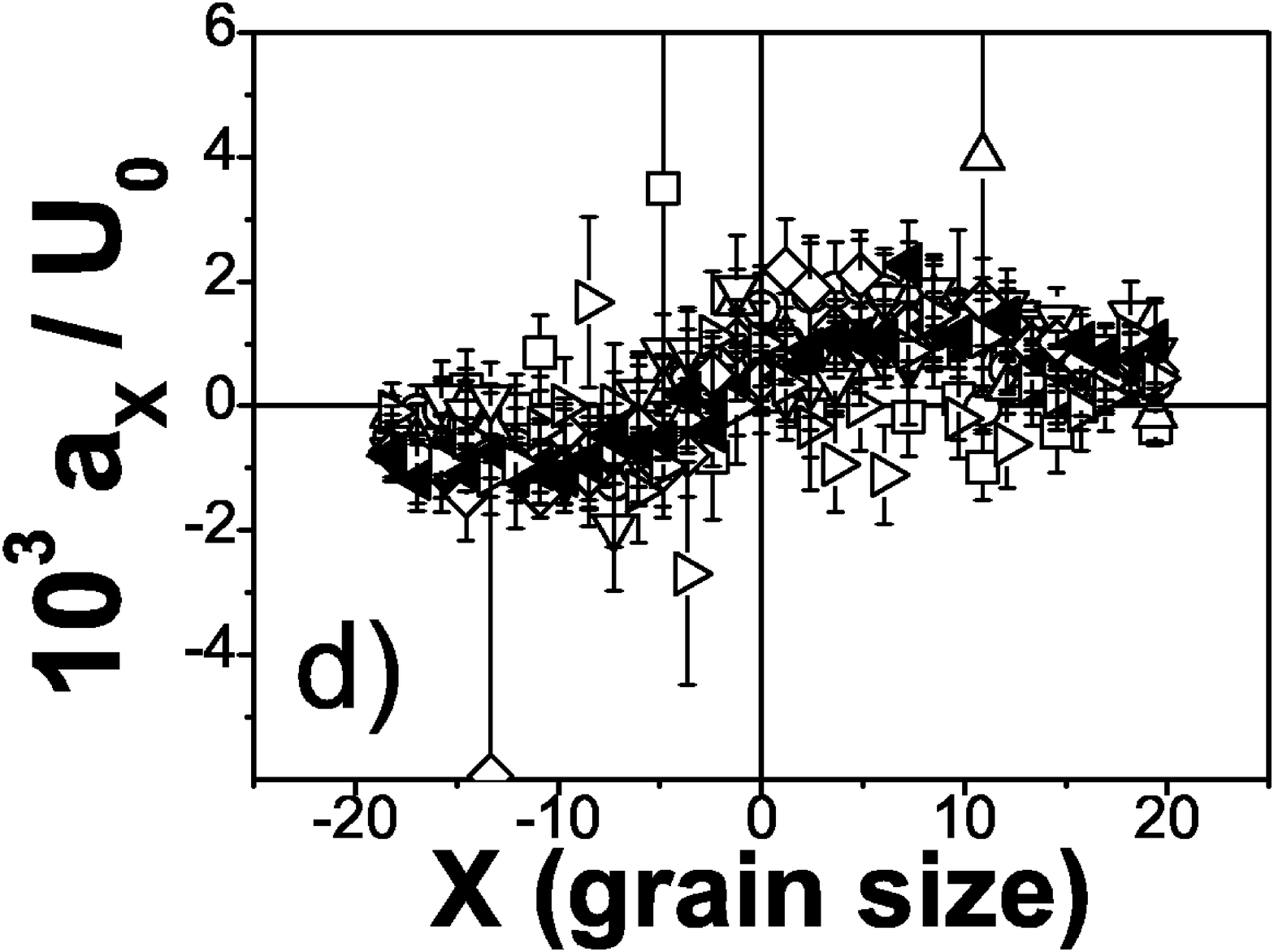,width=0.45\linewidth}
\end{center}
\caption{
a) and b) Mean rescaled response functions in polar coordinates
$r u_r$ and $r u_\theta$ for the upward motion $i=17$ ($n=9$) as a function
of $\theta$. This data collapse has been obtained with radial distances
ranging from $7.4$ to $21.8 d_2$, represented by different symbols -- e.g.
$\blacktriangledown$ for $r=15.8 d_2$.
c) and d) Vertical and horizontal components of the irreversible displacement
field $\protect\overrightarrow{a}$ for cycle $n=9$ in cartesian coordinates as
a function of $X$. The horizontal span corresponds to the whole cell width.
Here the data correspond to $Y$ between $14.5$ and $26.7 d_2$ -- e.g.
$Y=24.2 d_2$ is $\blacktriangleleft$. Rescaling factors are indicated on the
axis legends. The quantities $b_n$ and $\epsilon_n$ are the maxima of the
mean profiles a) and c) respectively.}
\label{repprofiles}
\end{figure*}

To compute the averaged displacement fields we first coarse grain individual
grain displacements in little cells of typical size $1.2 d_2 \times 1.2 d_2$
regulary located in the cartesian coordinates ($O,X,Y$) reference frame
-- see figure \ref{photoecrou}. We then use an ensemble average. To this end,
$16$ equivalent experiments were performed by fixing the initial compacity at a
value $\phi =0.760 \pm 0.005$ and using the same preparation method for each
packing: the initial grain configuration is randomly mixed \emph{in the
horizontal position} the packing lying on the bottom plane and then slowly
inclined at $\varphi$. We directly tested that packing fraction does not
vary by more than $0.5\%$ during each experiment. The final precision for
the mean displacement per cell is less than $2$ $\mu$m. Hence, we obtain for
each cycle $n$ the displacement fields $\overrightarrow{u}_{n}^{\uparrow }$
in response to the upward intruder motion and
$\overrightarrow{u}_{n}^{\downarrow }$ in
response to the subsequent downward motion. The irreversible or plastic
displacement field $\overrightarrow{a}_{n}$ is defined through the relation: 
$\overrightarrow{u}_{n}^{\downarrow }=-\overrightarrow{u}_{n}^{\uparrow }+
\overrightarrow{a}_{n}$.

On figure \ref{champdepl}, we show the ensemble average and coarse-grained
displacement field obtained after the second upward motion of the intruder
(cycle $n=2$, displacement $i=3$). We clearly notice that the granular motion
is not localized in the vicinity of the intruder and that this small
perturbation of only one third of a grain diameter produces a far field
effect. The presence of two displacement rolls is observed near the intruder.
They are located symmetrically on each side of the intruder but turn in
opposite directions. Besides this near field effect, the main response
principally occurs above the intruder with displacement vectors that tend
to align along the radial directions from the intruder. On figures
\ref{repprofiles}(a) and \ref{repprofiles}(b) we display a typical response
to the $n=9^{th}$ upward motion ($i=17$) in polar coordinates ($0,r,\theta $).
By defining $\overrightarrow{u}_{n}^{\uparrow}=
u_{n,r}\overrightarrow{e}_{r}\ \ + u_{n,\theta }\overrightarrow{e}_{\theta}$,
we display the rescaled quantities: $ru_{n,r}$ and $ru_{n,\theta }$ as a
function of the angle $\theta $. The collapse of the data onto the same
curve (within experimental uncertainties) at distances far enough from the
intruder ($r\geq 6$ $d_{2})$, shows that we can extract a \emph{dominant}
term for the far field displacement of the type: 
\begin{equation}
\overrightarrow{u}_{n}^{\uparrow }   \simeq
b_n U_0 d_2  \frac{f(\theta )}{r}    \overrightarrow{e}_{r},
\label{Formula2}
\end{equation}
The parameter $b_{n}$ is a dimensionless decreasing function of the cycle
number $n$ (see further) and $f(\theta )$ is a even function of $\theta $
with $f(0)=1$. Note that we could not observe any significant shape
variation of the function $f(\theta )$ with $n$. The preceding formula holds
for the zone above the intruder, i.e. $-90^{o}\leq \theta \leq 90^{o}$, as
the decay of the response below the intruder is much faster than $1/r$.
According to equation (\ref{Formula2}) the projection
$u_{n,Y}(X,Y)$ of $\overrightarrow{u}_{n}^{\uparrow }$ along
$\overrightarrow{e}_{Y}$ should be maximal in $X=0$ for a given vertical
distance $Y$ from the intruder, leading to
$u_{n,Y max}(Y) \simeq b_n U_0 d_2/Y$.
Simple integration of $u_{n,Y}$ along the $X$ axis shows that the quantity 
$b_n U_0 d_2$ is proportional to the effective area displaced by the
intruder on the horizontal line seen at a remote vertical distance $Y$.
Thus, in this interpretation, parameter $b_{n}$ can be seen as an `effective
transmission' parameter or a `displacement impedance' connecting the
externally driven motion of the intruder to the granular reorganizations in
the bulk. A good determination of $b_{n}$ can be inferred from average of
the experimental quantity $\beta_n = Y u_{n,Y max}/(U_{0}d_{2})$ over
values of $Y$ in between $6 d_{2}$ and $27 d_{2}$. 

\begin{figure}[t]
\begin{center}
\epsfig{file=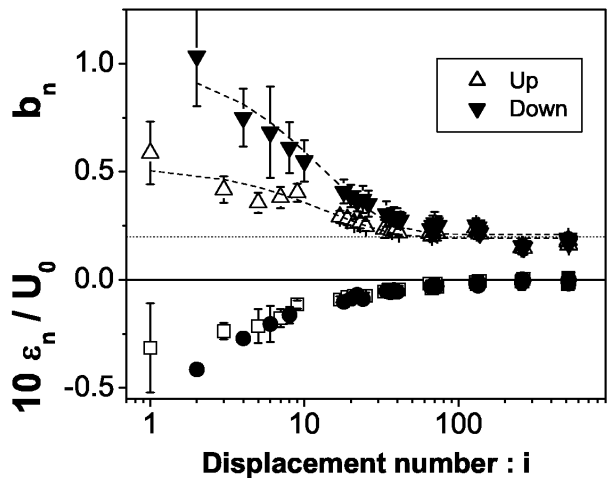,width=0.93\linewidth}
\end{center}
\caption{
Evolution of the plateau value $b_{n}$ (upper part) and the rescaled
irreversible component $\epsilon_n \equiv a_{n,max}/U_0$ (bottom part)
versus the displacement number $i$ or equivalently $n$ with
$n=int\left(\frac{i+1}{2}\right)$. The value $a_{n,max}$ is computed
as the mean of the maximum vertical component of the irreversible field
$\protect\overrightarrow{a}_n$. The definition of $b_n$ initially introduced
for upward motions of the intruder is extended to downward motions.}
\label{bn}
\end{figure}

Now we study the behavior of the irreversible field $\overrightarrow{a}_{n}$
for the transient part of the response. After few cycles its amplitude is
relatively small: one tenth or less of the displacement field magnitude. On
figures \ref{repprofiles}(c) and \ref{repprofiles}(d), we display the
cartesian coordinates of $\overrightarrow{a}_{n} =
a_{n,X}\overrightarrow{e}_{X} + a_{n,Y}\overrightarrow{e}_{Y}$,
for a cycle number $n=9$ and for different heights far from intruder level.
We observe a striking feature: the dominant part of the field has an
amplitude almost independent of the vertical distance and its influence
spans the whole width of the container ($27 d_2$ on each side). It can be
described as a quasi vertical columnar flow with a linear decay of its
amplitude out of the median axis, as to first order, the shape of the field
is triangular with a maximum value $a_{n,max}$ in $X=0$. The measurement error
bars prevents more precise determination of the field shape. We also notice
that the horizontal part of the field $a_{n,X}$ is not exactly zero and could
actually show a slight tendency to point outwards the median line but we are
at the limit of the signal over noise ratio to go further in the analysis.

\begin{figure}[t]
\begin{center}
\epsfig{file=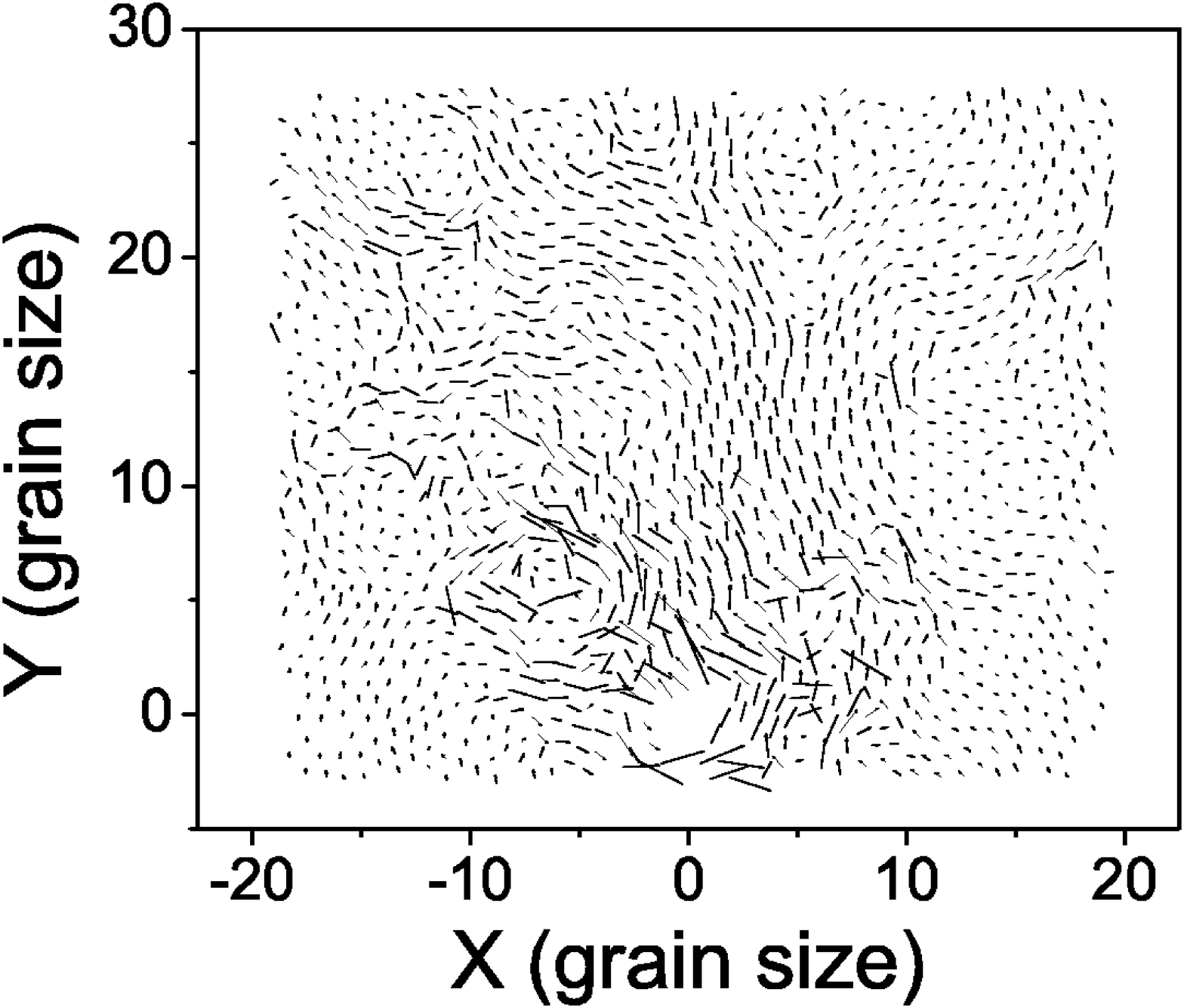,width=0.93\linewidth}
\end{center}
\caption{
Irreversible displacement field for a given experiment. The arrows
represent the cumulative displacement of each grain between the
images $i = 129$ and $i = 513$ (magnified by a factor of 20).}
\label{irrev}
\end{figure}

Both reversible and irreversible displacement fields evolve with the number of
cycles. On the upper part of figure \ref{bn}, we display the evolution of the
values $b_n$ as a function of the displacement number $i$. The error bar on
$b_n$ is a witness of the dispersion of the quantity $\beta _n$: it is larger
for the two first cycles where there is a departure from the scaling proposed
in (\ref{Formula2}). From figure \ref{bn} we observe a quasi exponential
relaxation to a limiting value $b_n=0.19 \pm 0.01$ which is $3$ times smaller
than the original one. This decrease of $b_n$ can be interpreted as a
progressive screening of the perturbation and is due to a local arching
effect that takes place at a distance of few grains around the intruder.
These observations seem to validate, at least in the limit of experimental
uncertainties, the interpretation of $b_{n}$\ as a displacement impedance.
The screening due to this arching effect progressively ceases to vary for a
number of cycles larger than few tens as evidenced by the steady saturation
of the $b_{n}$ values for $n\gtrsim 30$. In parallel (figure \ref{bn}, bottom
part), we monitor the mean maximal amplitude $a_{n,max}$ of the irreversible
component $a_{n,Y}$. We observe its decay to zero for $i\gtrsim 60$, i.e.
$n\gtrsim 30$ which corresponds to the onset of a quasi-reversible response
observed from the curve $b_{n}(i).$

In the quasi-reversible regime ranging from $n=30$ up to $n=262$ for the
longest experiments we did, we could not observe any sensible variation of
the ensemble average fields ($\overrightarrow{u}_{n}^{\uparrow }$ or
$\overrightarrow{a}_{n}$) and compacity. On the other hand, a closer look at 
\emph{each individual experimental realization} of the irreversible
displacement field, shows a very striking feature -- see figure \ref{irrev}.
The irreversible displacement field has radically changed its symmetry:
coherent streams and vortex-like structures appear in the whole packing.
What is not clear yet is whether this structuring will lead to a further
slower evolution of the packing or whether it corresponds to some
steady-state feature i.e. a remanent steady dissipation field. The long
range coherent structures found in many different situations are in our
opinion of a central importance since they could carry some universal
features characterizing the collective dissipation modes of condensed
jammed phases.

In conclusion, we propose the first experimental determination
of the `unjamming response' i.e. the reorganization field
due to a localized cyclic displacement experienced by a packing of
hard-grains under gravity. We use a small perturbation such as to
investigate intermediate values of deformations which are large compared to
the deformations at granular contacts but small compared with usual
experiments where a plastic deformation field is fully developed. The response
to an upward perturbation can be separated into three distinct parts. Far from
the intruder we observe a dominant radial displacement field which amplitude
scales as the inverse of the distance to the perturbation. Close to the
perturbation we observe displacement rolls on both sides as well as a vault
forming in the immediate surrounding of the intruder. In the part below the
intruder the displacement decays rapidly to zero. A local arching effect
progressively screens the perturbation as seen far from the intruder and is
accompanied by an irreversible field shaped as a quasi-vertical columnar
flow. This irreversible downward flow is too small to induce noticeable
changes of compacity but is sufficient to modify the subsequent mechanical
response. Thus the granular material can be seen as an auto-adaptative
material screening the exerted perturbation. Once this flow ceases, the
response is quasi reversible but on each realization, we still observe a
remanent irreversibility flow spanning the whole container and characterized
by a different symmetry: vortex-like structures and coherent long range
streams. These vortex-like structures remind strongly what was observed
recently in simulations for the non-affine components of the elastic field
in amorphous media \cite{Wittmer} and also for the particle displacement
fluctuations in a quasi statically sheared granular flow \cite{Radjai2}.
The question whether these features are linked to the 2D aspect of the
studied systems, and that 3D granular assemblies may behave differently
has to be investigated. In the future we plan to compare these experimental
data with theoretical predictions, e.g. elastic calculations such as those
computed in \cite{Wittmer} or \cite{Graner} in the context of 2D amorphous
media. At last, it would be interesting to monitor the change of the
unjamming response function with respect to initial compacity and/or to
texture parameters in association with force measurements. This systematic
study could bring some crucial informations on the nature of the
jamming/unjamming transition.


\centerline{\rule[0.1cm]{3cm}{1pt}}

We thank J.-L. Barrat, R.P. Behringer, R. Bouamrane, J.-C. G\'eminard,
J. Goddard, I. Goldhirsch, J.E.S. Socolar, A. Tanguy and J.P. Wittmer for
fruitful interactions and comments. E.K. also thanks J. Treiner for his kind
support. Finally, we are grateful to A. Poirier and L. Baouze for their
experimental contribution during their undergraduate training period.


\end{document}